\documentclass{moriond}

\def\be{\begin{equation}}
\def\ee{\end{equation}}
\def\bea{\begin{eqnarray}}
\def\eea{\end{eqnarray}}

\newcommand{\Photo}{\includegraphics[height=35mm]{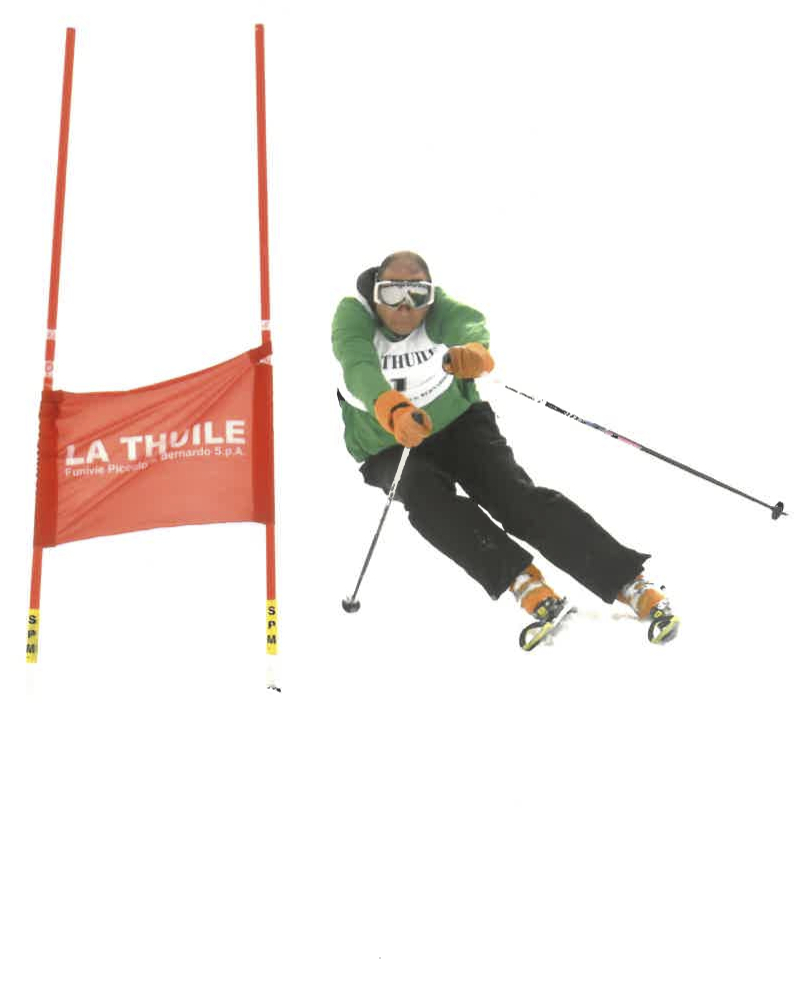}}

\begin{document}
\vspace*{4cm}
\title{Towards a minimal $SU(5)$ model}

\author{Ilja Dor\v{s}ner}

\address{University of Split, Faculty of Electrical Engineering, Mechanical Engineering and Naval Architecture (FESB)\\ 
Ru\dj era Bo\v skovi\' ca 32, 21000 Split, Croatia}

\maketitle\abstracts{I briefly outline the most prominent features of a novel $SU(5)$ unification model proposal. The particle content of the model comprises $5_H$, $24_H$, $35_H$, ${\overline{5}_F}_i$, ${10_F}_i$, $15_F$, $\overline{15}_F$, and $24_V$, where  subscripts $H$, $F$, and $V$ denote whether a given representation contains scalars, fermions, or gauge bosons, respectively, while $i=1,2,3$. The model employs all possible interaction terms, as allowed by the Lorentz group, the $SU(5)$ gauge symmetry, and the aforementioned particle content, to generate the Standard Model fermion masses through three different mechanisms. It also connects the neutrino mass generation mechanism to the experimentally observed mass disparity between the down-type quarks and charged leptons. The minimal structure of the model requires $24_H$ and $5_H$ not only to break $SU(5)$ and $SU(3) \times SU(2) \times U(1)$ gauge groups, respectively, but to accomplish one additional task each. The model furthermore predicts that neutrinos are strictly Majorana fields, that one neutrino is purely a massless particle, and that neutrino masses are of normal ordering. The experimental bound on the $p \rightarrow \pi^0 e^+$ lifetime limit, in conjunction with the prediction of the model for the gauge mediated proton decays, implies that there are four new scalar multiplets at or below a $120$\,TeV mass scale if these multiplets are mass degenerate. If these multiplets are not mass degenerate, the quoted limit then applies, for all practical purposes, to the geometric mean of their masses. The scalars in question transform as $(1,3,0)$, $(8,1,0)$, $(\overline{3},3,-2/3)$, and $(\overline{6},2,1/6)$ under the Standard Model gauge group $SU(3) \times SU(2) \times U(1)$ with calculable couplings to the Standard Model fields.}

\section{A novel $SU(5)$ model proposal}
I present, in what follows, a minimal realistic $SU(5)$ model~\cite{Dorsner:2019vgf} and a phenomenological study~\cite{Dorsner:2021qwg} of the viability of its parameter space. The model under consideration is a simple extension of the original Georgi-Glashow proposal~\cite{Georgi:1974sy}. 

The particle content of the Georgi-Glashow model comprises $5_H$, $24_H$, ${\overline{5}_F}_i$, ${10_F}_i$, and $24_V$, where subscripts $H$, $F$, and $V$ denote whether a given representation contains scalars, fermions, or gauge bosons, respectively, while $i(=1,2,3)$ represents a generation index. (Note that the $SU(5)$ representations are simply identified through their dimensionality. I will use the same approach when denoting the Standard Model gauge group $SU(3) \times SU(2) \times U(1)$ multiplets.) The novel proposal in question~\cite{Dorsner:2019vgf} extends the Georgi-Glashow particle content with one 35-dimensional scalar representation, i.e., ${35}_H$, and one set of vector-like fermions in 15-dimensional representation comprising $15_F$ and $\overline{15}_F$. The particle content of the model is presented in Table~\ref{table:content}, where I also specify the decomposition of the $SU(5)$ representations into the Standard Model multiplets and introduce associated nomenclature.  
\begin{table}[h]
\caption{The particle content of the novel proposal at both the $SU(5)$ and the Standard Model levels. Subscripts $H$ and $F$ denote scalar and fermion representations, respectively, while $i(=1,2,3)$ is a generation index.}
\vspace{0.4cm}
\begin{center}
\begin{tabular}{| c | c | c | c |}
\hline
$SU(5)$ & $SU(3)\times SU(2) \times U(1)$  & $SU(5)$ & $SU(3)\times SU(2) \times U(1)$ \\\hline
 & $\Lambda_1\left(1,2,\frac{1}{2}\right)$ & & $L_i\left(1,2,-\frac{1}{2}\right)$\\
\raisebox{2ex}[0pt]{$5_H \equiv \Lambda$} & $\Lambda_3\left(3,1,-\frac{1}{3}\right)$  & \raisebox{2ex}[0pt]{${\overline{5}_F}_i \equiv F_i$} & $d_i^c\left(\overline{3},1,\frac{1}{3}\right)$\\ 
\hline
 & $\phi_0 \left(1,1,0\right)$ & & $Q_i\left(3,2,\frac{1}{6}\right)$\\
 & $\phi_1 \left(1,3,0\right)$ & ${10_F}_i \equiv T_i$ & $u_i^c\left(\overline{3},1,-\frac{2}{3}\right)$\\
$24_H \equiv \phi$ & $\phi_3\left(3,2,-\frac{5}{6}\right)$ &  & $e_i^c \left(1,1,1\right)$\\\cline{3-4}
 & $\phi_{\overline{3}} \left(\overline{3},2,\frac{5}{6}\right)$ &  & $\Sigma_1(1,3,1)$\\
  & $\phi_8 \left(8,1,0\right)$ & $15_F \equiv \Sigma$ & $\Sigma_3\left(3,2,\frac{1}{6}\right)$\\\cline{1-2}
  & $\Phi_1  \left(1,4,-\frac{3}{2}\right)$ & & $\Sigma_6\left(6,1,-\frac{2}{3}\right)$\\\cline{3-4}
  & $\Phi_3  \left(\overline{3},3,-\frac{2}{3}\right)$ & & $\overline{\Sigma}_1\left(1,3,-1\right)$\\
 \raisebox{2ex}[0pt]{$35_H \equiv \Phi$} & $\Phi_6  \left(\overline{6},2,\frac{1}{6}\right)$ & $\overline{15}_F \equiv \overline{\Sigma}$ & $\overline{\Sigma}_3\left(\overline{3},2,-\frac{1}{6}\right)$\\
 & $\Phi_{10}  \left(\overline{10},1,1\right)$ &  & $\overline{\Sigma}_6\left(\overline{6},1,\frac{2}{3}\right)$\\
\hline  
\end{tabular}
\end{center}
\label{table:content}
\end{table}

The fermionic interactions of the model are governed by the following lagrangian
\begin{eqnarray}
\mathcal{L} \supset &&\{Y^u_{ij} T^{\alpha\beta}_iT^{\gamma\delta}_j\Lambda^\rho \epsilon_{\alpha\beta\gamma\delta\rho} 
+Y^d_{ij} T^{\alpha\beta}_iF_{\alpha j} \Lambda^{\ast}_\beta 
+Y^{a}_{i} \Sigma^{\alpha\beta}F_{\alpha i}  \Lambda^{\ast}_\beta
+Y^{b}_{i} \overline{\Sigma}_{\beta\gamma}F_{\alpha i} \Phi^{*\alpha\beta\gamma} 
\nonumber \\ 
 &&
+Y^{c}_{i} T^{\alpha\beta}_i  \overline{\Sigma}_{\beta\gamma}\phi^\gamma_\alpha+\mathrm{h.c.}\}+M_{\Sigma} \overline{\Sigma}_{\alpha\beta} \Sigma^{\alpha\beta}
+y \overline{\Sigma}_{\alpha\beta} \Sigma^{\beta\gamma} \phi_\gamma^\alpha,
\label{eq:lagrangian} 
\end{eqnarray}
where the Yukawa matrix elements are $Y^u_{ij} \equiv Y^u_{ji}$, $Y^d_{ij} = Y^{d*}_{ij} \equiv \delta_{ij} Y^d_{i}$, $Y^{a}_{i}$, $Y^{b}_{i}$, $Y^{c}_{i}$, and $y$ with $i,j=1,2,3$. The model has nineteen real Yukawa couplings and fourteen phases, all in all. There is accordingly no other $SU(5)$ model in the literature that successfully incorporates neutrino mass generation mechanism that has fewer parameters than the one under consideration.

The role of the scalar field $\phi_0 \left(1,1,0\right) \in {24}_H$ in the Georgi-Glashow proposal is solely to break $SU(5)$ down to $SU(3) \times SU(2) \times U(1)$ whereas the scalar multplet $\Lambda_1(1,2,1/2) \in 5_H$ breaks $SU(3) \times SU(2) \times U(1)$ down to $SU(3) \times U(1)_\mathrm{em}$ and, in the process, generates masses of the charged fermions. In this proposal, on the other hand, both $\phi_0 (1,1,0)$ and $\Lambda_1(1,2,1/2)$ have additional tasks to accomplish beside the ones in the Georgi-Glashow model. Namely, $\phi_0 (1,1,0)$ is instrumental in generating the experimentally observed mismatch between the down-type quark masses and the charged lepton masses whereas $\Lambda_1(1,2,1/2)$ is essential in producing neutrino masses via a one-loop level mechanism. The fact that both $\phi_0(1,1,0)$ and $\Lambda_1(1,2,1/2)$ have additional roles to play, when compared to the original Georgi-Glashow model, attests to the simplicity of this novel proposal~\cite{Dorsner:2019vgf}.

\begin{figure}[t!]
\centering
\includegraphics[width=0.47\textwidth]{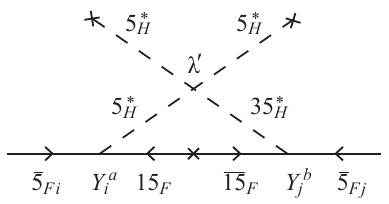}
\includegraphics[width=0.47\textwidth]{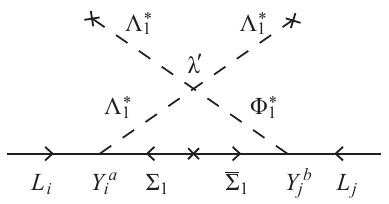}
\caption{The Feynman diagrams of the leading order contribution towards the Majorana neutrino masses at the $SU(5)$ (left panel) and the Standard Model (right panel) levels.}
\label{fig:diagram}
\end{figure}

The model predicts that the neutrino masses are purely of the Majorana nature. The leading order contribution is generated at the one-loop level via the $d=5$ operator. The relevant Feynman diagrams, both at the $SU(5)$~\cite{Dorsner:2019vgf} and the Standard Model~\cite{Babu:2009aq,Bambhaniya:2013yca} levels, are shown in Fig.~\ref{fig:diagram}. Clearly, $\Lambda_1(1,2,1/2)$ and its vacuum expectation value $\langle 5_H \rangle \equiv \langle\Lambda_1(1,2,1/2)\rangle=(0 \quad 0 \quad 0 \quad 0 \quad v_{5})^T$ are indispensable to the neutrino mass generation. (The relevant contraction that yields the quartic scalar coupling vertex in Fig.~\ref{fig:diagram} is $\lambda^\prime \Lambda^\alpha\Lambda^\beta\Lambda^\gamma \Phi_{\alpha\beta\gamma}$.) 

A presence of the vector-like fermions in $15_F$ and $\overline{15}_F$ induces experimentally observed mismatch between the masses of the charged leptons and the down-type quarks. The mismatch itself is due to a physical mixing between the vector-like fermions and fermions in ${10_{F}}_i$. (The effect of this type of mixing on the charged fermion masses has already been studied within the context of a supersymmetric $SU(5)$ framework~\cite{Oshimo:2009ia}.) Namely, since the quark doublets $Q_i \in {10_F}_i$ and a multiplet $\Sigma_3 \in 15_F$ transform in the same way under the Standard Model gauge group, as can be seen from Table~\ref{table:content}, these states interact at the $SU(5)$ symmetry breaking level, where the relevant mixing originates from the first term in the second line of Eq.~\ref{eq:lagrangian} and explicitly reads 
\begin{equation}
\mathcal{L} \supset \frac{1}{4}\sqrt{\frac{10}{3}}v_{24} Y^c_i\; Q_i\overline{\Sigma}_3.
\end{equation}
Here, $v_{24}$ is defined via the vacuum expectation value of $24_H$ as $\langle 24_H \rangle \equiv \langle\phi_0(1,1,0)\rangle=v_{24}/\sqrt{15} \ \textrm{diag}(1,1,1,-3/2,-3/2)$. One can see that it is the vacuum expectation value of $24_H$ that plays an instrumental role in generating the observed mismatch between the masses of the charged leptons and the down-type quarks.

The mass matrices for the up-type quarks, down-type quarks, charged leptons, and neutrinos, in this model, are
\begin{eqnarray}
(M_U)_{ij}&=& 4 \,v_5\, (Y^{u}_{ij}+Y^{u}_{ji}),\\
(M_D)_{ij}&=& v_5 \left( \,\delta_{ij} Y^d_{i} + \delta^\prime\; Y^c_i Y^a_j \, \right),\\
(M_E)_{ij}&=&v_5\, \delta_{ij} Y^d_{i},\\
(M_N)_{ij}
&\approx& \frac{\lambda^{\prime}v_5^2}{8 \pi^2} \frac{M_{\Sigma_1}}{M^2_{\Sigma_1}-M^2_{\Phi_1}}
\ln \left( \frac{M^2_{\Sigma_1}}{M^2_{\Phi_1}} \right) (Y^a_iY^b_j+Y^b_iY^a_j)  = m_0 (Y^a_iY^b_j+Y^b_iY^a_j),
\label{eq:massnu}
\end{eqnarray}
where $\delta^\prime \equiv (\sqrt{10/3} v_{24})/(4 M_{\Sigma_3})$. The mismatch between the down-type quark and charged lepton masses clearly originates from a single rank-one matrix with elements proportional to $Y^c_i Y^a_j$ products. This, again, attests to the simplicity of this $SU(5)$ proposal. It should also be noted that $M_D$ and $M_N$ share $Y^a$ column matrix. The generation of the mismatch between the down-type quark and charged lepton masses is thus inextricably connected to the generation of the neutrino masses. This link is behind one of the predictions of the model that neutrino masses are strictly of the normal ordering nature. It is furthermore clear from Eq.~\ref{eq:massnu} that $M_N$ is constructed in the most minimal way imaginable out of two rank-one matrices with elements $Y^a_iY^b_j$ and $Y^b_iY^a_j$. This guarantees that one neutrino is massless which is one of the main predictions of the model. Note also that the model simultaneously generates viable masses for the Standard Model fermions through the usual vacuum expectation value mechanism, the one-loop level mechanism, and the mixing between chiral fields and vector-like states~\cite{Witten:1979nr}. 

A parameter $m_0$ of Eq.~\ref{eq:massnu} sets the neutrino mass scale through its dependence on {\it a priori} unknown parameters $M_{\Phi_1}$, $M_{\Sigma_1}$, and $\lambda^{\prime}$. There is thus a constrain on the available parameter space of the model in the $M_{\Phi_1}$-$M_{\Sigma_1}$ plane that originates solely from a need to generate sufficiently large $m_0$ parameter. I will, in order to enlarge $m_0$ as much as possible, use $|\lambda^{\prime}|=1$ in the numerical analysis. (Strictly speaking, $Y^a$ and/or $Y^b$ can always be redefined as to make $\lambda^{\prime}$ real and positive.)

This completes a brief introduction of the model.

\section{Parameter space analysis}

\begin{figure}[th!]
\centering
\includegraphics[width=0.48\textwidth]{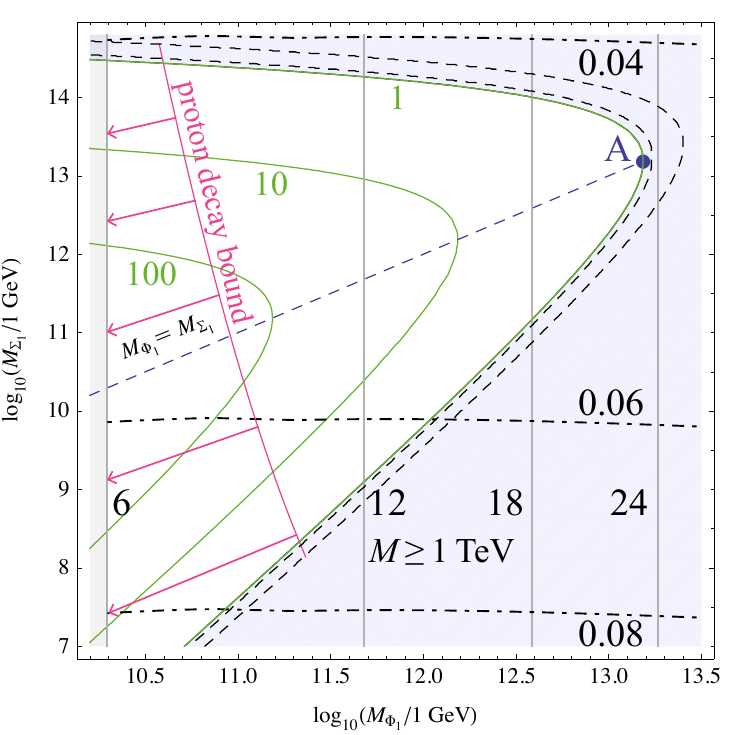}
\includegraphics[width=0.5\textwidth]{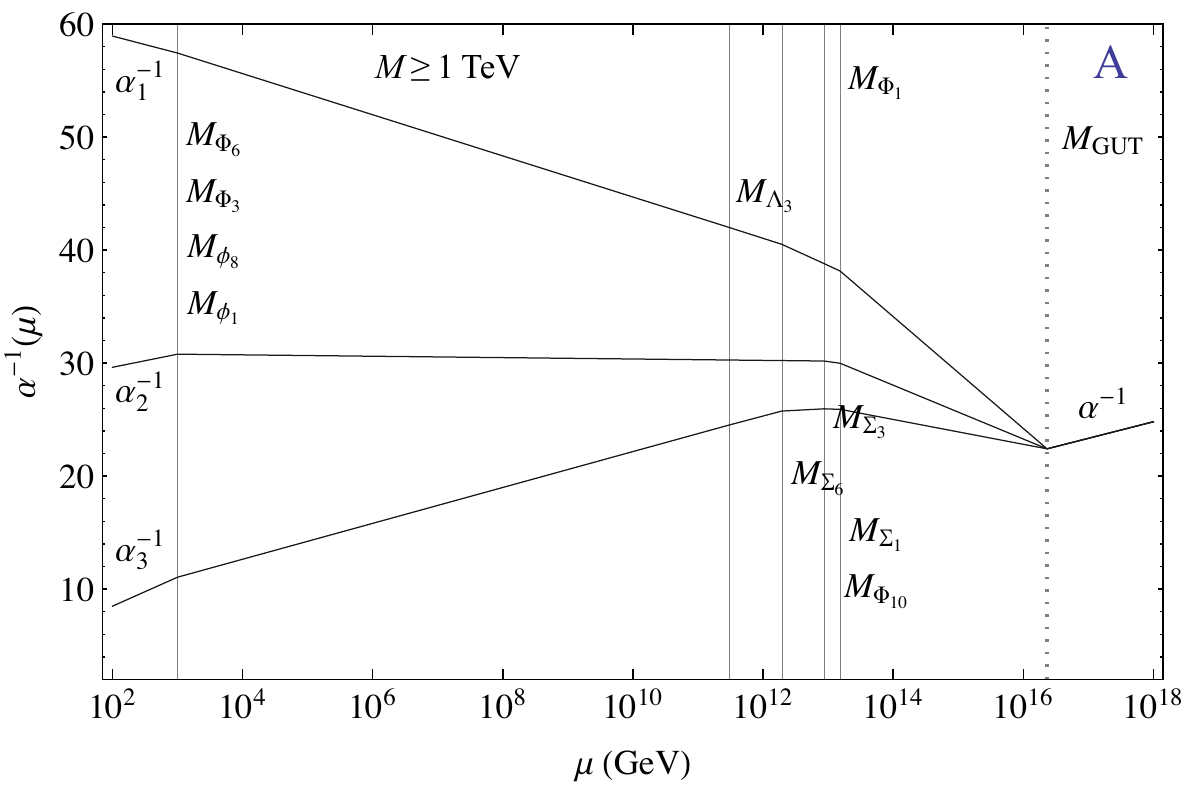}
\includegraphics[width=0.48\textwidth]{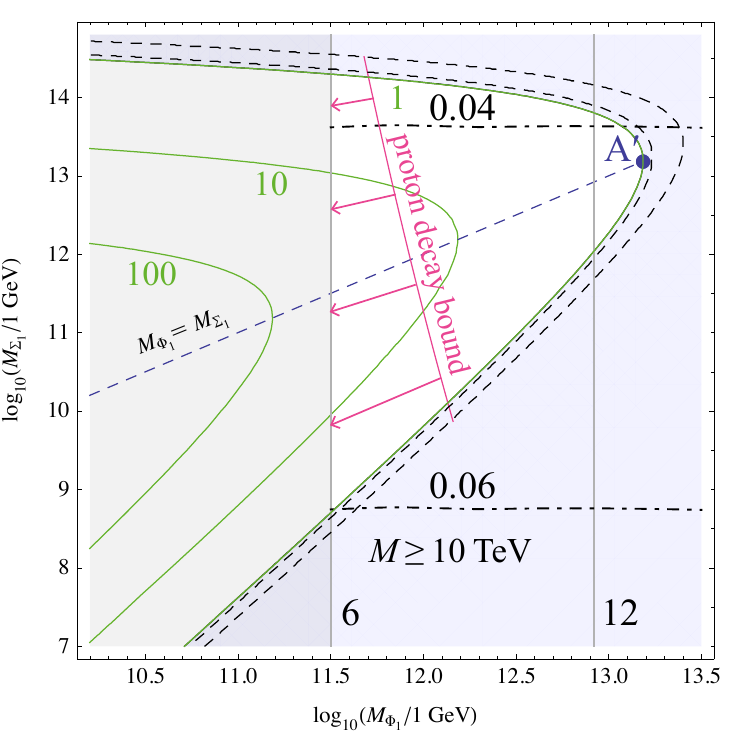}
\includegraphics[width=0.5\textwidth]{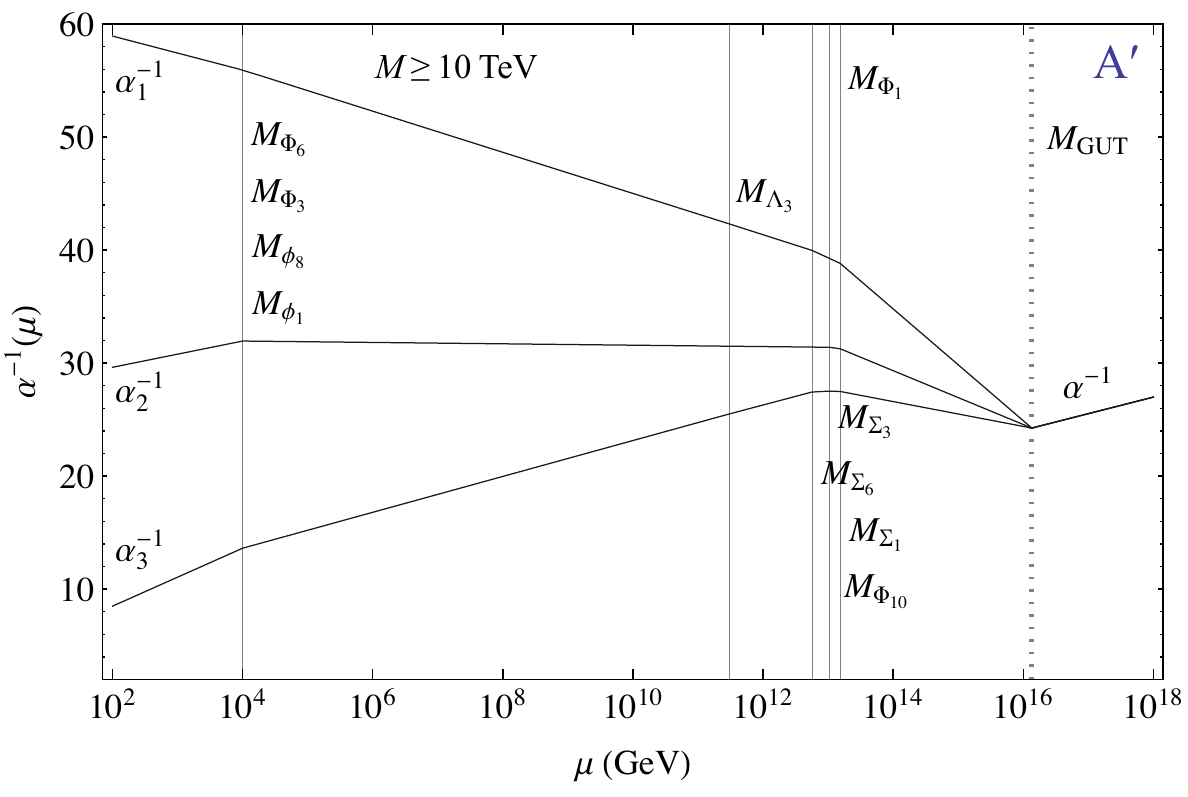}
\includegraphics[width=0.48\textwidth]{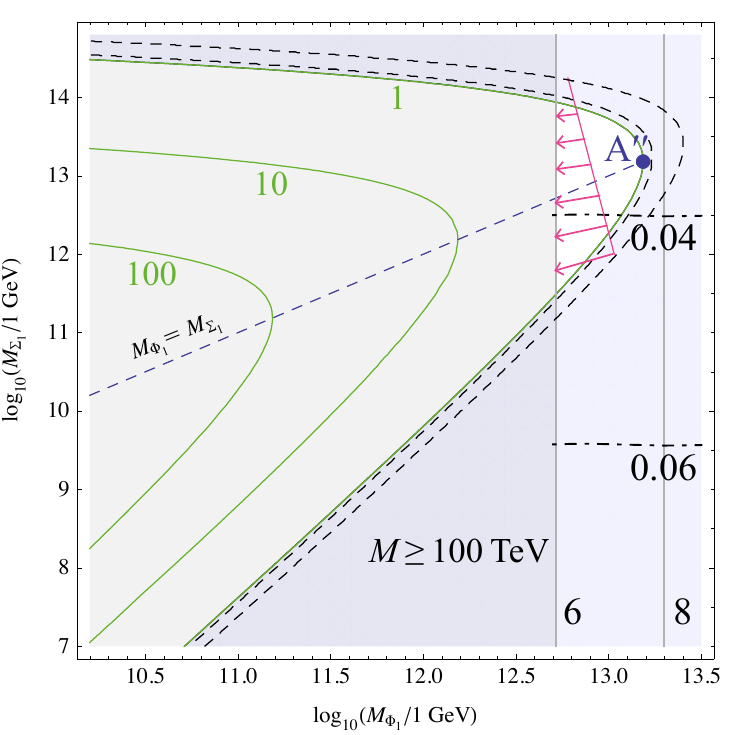}
\includegraphics[width=0.5\textwidth]{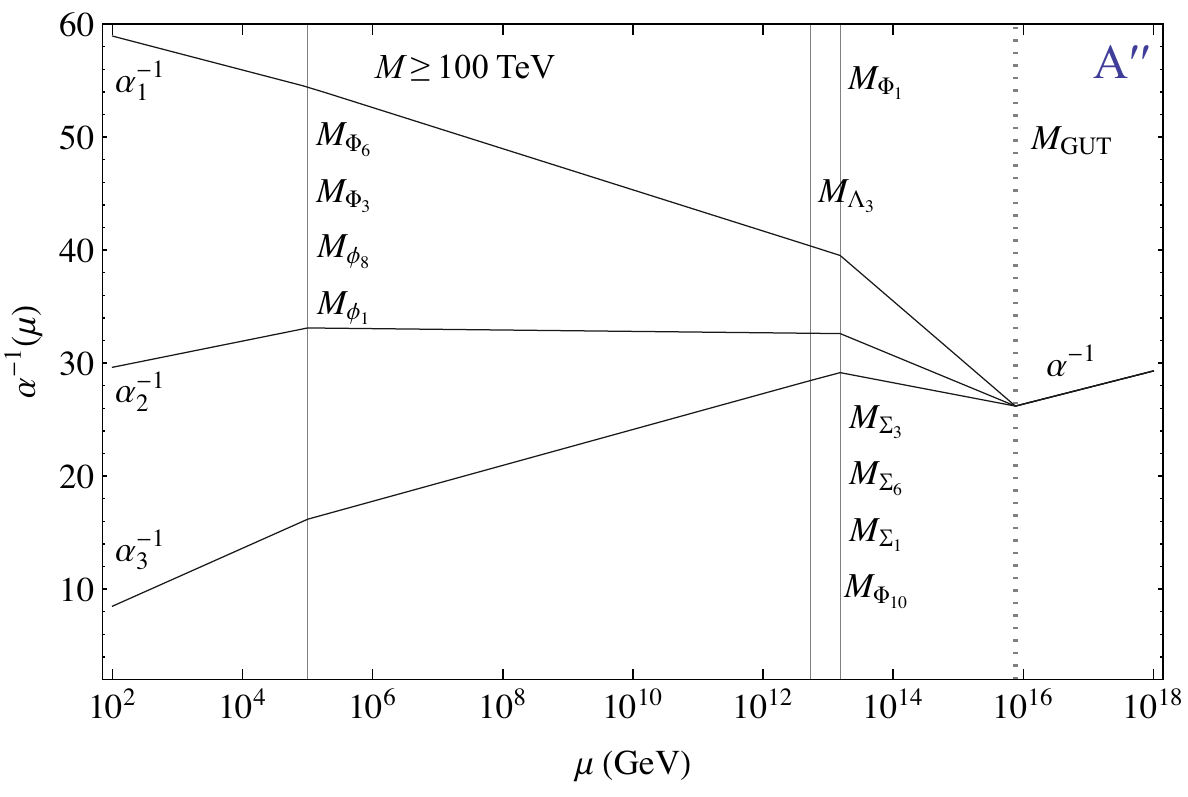}
\caption{Experimentally viable parameter space of the model (left panels) and the gauge coupling unification for the unification points A, A$^\prime$, and A$^{\prime\prime}$ (right panels) when $M \geq 1, 10, 100$\,TeV, as indicated. The contours of constant $M_\mathrm{GUT}$ are given in units of $10^{15}$\,GeV and appear as vertical solid lines while the contours of the gauge coupling $\alpha_\mathrm{GUT}$, at $M_\mathrm{GUT}$ scale, are given as dot-dashed lines that run horizontally. Contours of constant $m_0$ are shown as green solid curves in units of $\sqrt{\Delta m^2_{31}}/2$, where $\Delta m^2_{31}$ is the relevant neutrino mass square difference.}
\label{fig:MASTER}
\end{figure}

A one-loop level gauge coupling unification analysis that aims at finding the largest possible value of the gauge coupling unification scale $M_\mathrm{GUT}$ reveals~\cite{Dorsner:2019vgf,Dorsner:2021qwg} that four scalar multiplets need to be light. These are $\phi_1(1,3,0) \in 24_H$, $\phi_8(8,1,0) \in 24_H$, $\Phi_3(\overline{3},3,-2/3) \in 35_H$, and $\Phi_6(\overline{6},2,1/6) \in 35_H$. On the other hand, the multiplets in $15_F$ and $\overline{15}_F$ want to be mass degenerate. If a lower limit on the mass(es) of the new physics state(s) is introduced through a mass variable $M=\min(M_J)$, where $J=\Phi_1, \Phi_3, \Phi_6, \Phi_{10}, \Sigma_1, \Sigma_3, \Sigma_6, \phi_1, \phi_8, \Lambda_3$, the corresponding parameter space of the model that maximises $M_\mathrm{GUT}$ is as shown in the left panels of Fig.~\ref{fig:MASTER} for $M=1$\,TeV, $M=10$\,TeV, and $M=100$\,TeV. The viable parameter space is bounded from the left by the proton decay curve and from the right by the outermost dashed curve. The outermost dashed curve delineates the region after which it is not possible to address phenomenologically viable neutrino mass scale with perturbative couplings. (Proton decay induced by $\Lambda_3$ exchange stipulates that $M_{\Lambda_3} \geq 3 \times 10^{11}$\,GeV in order to have agreement with the current experimental limits on $p \rightarrow K^+ \bar{\nu}$~\cite{Dorsner:2012uz}. This is one of the self-consistency constraints that needs to be imposed within the gauge coupling unification analysis.)

Contours of constant $m_0$ of Eq.~\ref{eq:massnu} are shown as green solid curves and demonstrate that the neutrino mass scale can vary within three orders of magnitude in the most conservative scenario when $M=1\,\mathrm{TeV}$. This is another nice feature of the model. The $m_0$ contours are in units of $\sqrt{\Delta m^2_{31}}/2$, where $\Delta m^2_{31} =(2.515 \pm 0.028) \times 10^{-3}$\,eV$^2$~\cite{Esteban:2020cvm} is the relevant neutrino mass square difference.

The contours of constant $M_\mathrm{GUT}$ are given in units of $10^{15}$\,GeV and appear as vertical solid lines while the contours of the gauge coupling $\alpha_\mathrm{GUT}$, at $M_\mathrm{GUT}$ scale, are given as dot-dashed lines that run horizontally in the left panels of Fig.~\ref{fig:MASTER}. (Note that the gauge coupling unification study, the associated numerical fit of fermion masses, and the proton decay analysis have all been performed whenever $M_\mathrm{GUT}$ has exceeded $6 \times 10^{15}$\,GeV to expedite the process, as indicated in the left panels of Fig.~\ref{fig:MASTER} with vertically shaded strips.)

The proton decay bound in Fig.~\ref{fig:MASTER} is generated by the experimental limit on the partial lifetime for the $p \rightarrow \pi^0 e^+$ process that is currently at $\tau^\mathrm{exp}_{p\to \pi^0 e^+} > 2.4 \times 10^{34}$\,years, as given by the Super-Kamiokande Collaboration~\cite{Takenaka:2020vqy}. An improvement of the current $p \rightarrow \pi^0 e^+$ lifetime limit by a factor of $2$, $15$, and $96$ would completely rule out the $M = 100$\,TeV, $M = 10$\,TeV, and $M = 1$\,TeV scenarios, respectively. The last viable point to be eliminated by the aforementioned improvement, in all three left panels of Fig.~\ref{fig:MASTER}, is $(M_{\Phi_1},M_{\Sigma_1})=(10^{13.2}\,\mathrm{GeV}, 10^{13.6}\,\mathrm{GeV})$. This is to be expected since  $\alpha_\mathrm{GUT}$ grows with a decrease in the $\Sigma_1$ mass for a fixed value of $M_{\Phi_1}$, whereas $M_\mathrm{GUT}$ remains constant.

Note that the numerical fit of fermion masses explicitly yields all unitary transformations and Yukawa couplings except for the phases associated with the up-type quark sector. The couplings and, accordingly, interactions of scalar multiplets $\phi_1(1,3,0)$, $\phi_8(8,1,0)$, $\Phi_3(\overline{3},3,-2/3)$, and $\Phi_6(\overline{6},2,1/6)$ are thus fully calculable. An analysis of the decay modes and associated lifetimes of these scalars as well as a full-fledged study of the proton decay signatures via the gauge boson and scalar leptoquark mediations is left for future publications.

\section{Conclusion}

I present a phenomenological study of the viable parameter space of the most minimal realistic $SU(5)$ model to date. The structure of the model is built entirely out of the fields residing in the first five lowest lying representations in terms of dimensionality that transform non-trivially under the $SU(5)$ gauge group. These representations are $5_H$, $24_H$, $35_H$, ${\overline{5}_F}_i$, ${10_F}_i$, $15_F$, $\overline{15}_F$, and $24_V$, where  subscripts $H$, $F$, and $V$ denote whether a given representation contains scalars, fermions, or gauge bosons, respectively, while $i=1,2,3$. The Yukawa couplings are $Y^u_{ij}=Y^u_{ji}$, $Y^d_{ij}=Y^{d*}_{ij}=\delta_{ij} Y^d_{i}$, $Y^{a}_{i}$, $Y^{b}_{i}$, $Y^{c}_{i}$, and $y$, where $i,j=1,2,3$. The model has nineteen real parameters and fourteen phases, all in all, to address experimental observables of the Standard Model fermions and accomplishes that via simultaneous use of three different mass generation mechanisms. It inextricably links the origin of the neutrino mass to the experimentally observed difference between the down-type quark and charged lepton masses. The main predictions of the model are that $(a)$ the neutrinos are Majorana particles, $(b)$ one neutrino is massless, $(c)$ the neutrinos have normal mass ordering, and $(d)$ there are four new scalar multiplets at or below a $120$\,TeV mass scale if they are mass degenerate. An improvement of the current $p \rightarrow \pi^0 e^+$ lifetime limit by a factor of $2$, $15$, and $96$ would require these four scalar multiplets to reside at or below the $100$\,TeV, $10$\,TeV, and $1$\,TeV mass scales, respectively, under the assumption of the multiplet mass degeneracy. If these multiplets are not mass degenerate, the quoted limits then apply, for all practical purposes, to the geometric mean of their masses. The scalar multiplets in question transform as $(1,3,0)$, $(8,1,0)$, $(\overline{3},3,-2/3)$, and $(\overline{6},2,1/6)$ under the Standard Model gauge group.

\section*{References}

\end{document}